\documentclass[aip,pof,amsmath,amssymb,reprint]{revtex4-1}

\usepackage{graphicx}
\usepackage{dcolumn}
\usepackage{bm}

\usepackage{color,soul}
\usepackage{epstopdf}

\begin{document}


\title{Flow and dispersion in anisotropic porous media: a Lattice-Boltzmann study} 



\author{D. Maggiolo}
\email[]{dario.maggiolo@unipd.it}

\author{F. Picano}
\email[]{francesco.picano@unipd.it}

\author{M. Guarnieri}
\email[]{massimo.guarnieri@unipd.it}
\affiliation{Department of Industrial Engineering, University of Padova, via Gradenigo 
6/A - 35131 Padua (PD), Italy}


\date{\today}

\begin{abstract}
Given their capability of spreading active chemical species and collecting electricity, 
porous media made of carbon fibers are extensively used as diffusion layers in energy 
storage systems, such as redox flow batteries. In spite of this, the dispersion dynamics 
of species inside porous media is still not well understood and often lends itself to different
interpretations. Actually, the microscopic design of efficient porous media which can
potentially and effectively improve the performances of flow batteries, is a still open challenge.
The present study aims to investigate the effect of fibrous media micro-structure on dispersion,
in particular the effect of fiber orientation on drag and dispersion dynamics. Several
Lattice-Boltzmann simulations of {flows through} differently-oriented fibrous media coupled with
Lagrangian simulations of particle tracers have been performed. Results show
that orienting fibers preferentially along the streamwise direction minimizes the drag and maximizes 
the dispersion, which is the most desirable condition for diffusion layers in flow batteries applications.
{\em This article is currently in press on Physics of Fluid journal.} 
\end{abstract}


\maketitle 




\section{\label{sec1} Introduction}

During the last years, the interest in the utilization of porous media composed of fibers has been considerably increased, especially for energy conversion applications~\cite{arico,zhai}. 
For instance, carbon papers and carbon felt are by now widely used as gas diffusion layers of fuel cells. But the rapid rise of decarbonized green energy demand does not limit the application of such materials to fuel cells. 
Flow batteries have recently been perceived as one of the most promising technologies for electrochemical energy storage. Even though flow batteries are known since the late {1980s}, it is only during recent years that the scientific community has focused on improving their performance~\cite{alotto,weber}. A cell of a flow battery is composed by two porous media fibrous electrodes. The inner surfaces of the porous media act as active site where electrochemical reduction and oxidation reactions of the electrolytes occur. 
Both half-cells are supplied with the electrolyte solutions which are stored in external tanks and circulated by pumps to keep on the reactions. One limitation to the peak performance of flow batteries consists of the too slow electrolyte transport in the electrodes~\cite{alotto}. The fluid dynamic optimization of the porous medium which provides both the electrochemical active surfaces and the mixing volume of the chemical species is one of the main technological issues {to be dealt with}~\cite{weber,bortolin}.

In fact, the slow dispersion process of species in water represents a
bottleneck for the peak performance of flow batteries. Specifically, the mass diffusion coefficients of the species in water, $D\approx 10^{-10} \ m^2/s$, are about 10000 smaller {then} the water kinematic viscosity, $\nu=\mu/\rho=10^{-6}\ m^2/s$, indicating that the mass diffusion is 10000 times slower than the momentum transport. 
Enhancing this diffusivity can produce a dramatic increase in the cell performance.
 A proper designed geometry of a non-isotropic porous medium can enhance this effective mass transport while minimizing the drag, thus improving and optimizing the batteries performances.\cite{weber}. The present study deals with such analyses, by means of a Lattice Boltzmann model and a Lagrangian Particle Tracking algorithm.

Even if the influence of medium porosity on the flow drag has been largely studied~\cite{whitaker-drag,jackson-drag}, the impact of its microscopic design on the combined mixing/transport mechanisms and drag is still not well assessed~\cite{kang,icardi}. In fact, even though the anomalous (i.e.\ non-Fickian) behavior of dispersion in porous media has been widely investigated~\cite{neuman,whitaker-book,bruderer}, it is not clear to what extent the micro-structure of the medium can impact macroscopic dispersion phenomena. Local heterogeneities at various scales have been considered capable of generating such anomalous behavior~\cite{neuman2009,berkowitz,koch}. Berkowitz and Sher~\cite{berkowitz} claimed that a wide distribution of delay times limiting the transport in porous media results in non-Fickian dispersion which cannot be represented by an equation including a time-dependent dispersion coefficient. Instead, the authors highlighted that all the time evolution of motion must be taken into account, and that the macroscopic advection-dispersion equation (ADE) must be non-local in time.

Whitaker~\cite{whitaker-book} identified different fluid-dynamic variables responsible for the dispersion by means of the Volume Averaging technique. This analysis revealed the presence of different terms in the averaged ADE which act as sources of dispersion and convection. 
Nevertheless, the volume averaging technique is not sufficient to predict the dispersion behavior in a general way, since the evaluation of the effective dispersion tensor is limited by some constraints. In fact, in practical applications the value of the effective dispersion tensor may be significantly different than expected, since it depends on the unconditioned statistics of hydraulic permeabilities of the porous medium~\cite{whitaker-disp}.

Several authors agree that dispersion should tend to the standard Fickian dispersion at a certain temporal or length scale for which all the hypothesis of the central limit theorem are satisfied, i.e.\ when the particle motion is no more correlated~\cite{moroni,wang}. Such transient anomalous behavior has been recently recognized in a variety of physical-chemical and socio-economical systems, which can also present non-Gaussian yet Fickian dispersion behaviors~\cite{wang}. However, the aforementioned time or length scales strongly depend on the medium structure and, thus, they are not easy to determine a priori.

More recently, other causes have been identified as responsible of the anomalous dispersion, such as the presence of three-dimensional vortices~\cite{cardenas}, particle jumps ~\cite{meerschaert} and different mechanisms of dispersion acting on subgroups of particles~\cite{castiglione}. Castiglione et al.~\cite{castiglione} suggested that two mechanisms of dispersion (i.e.\ a weak anomalous dispersion and a strong anomalous dispersion associated to ballistic motion) can give rise to transient anomalous dispersion in several systems. The authors underlined that even though it is not particularly difficult to build up probabilistic models exhibiting anomalous dispersion, understanding anomalous dispersion in nontrivial systems, such as porous media, is much more difficult~\cite{castiglione}.

A review of the literature about anomalous dispersion revealed that this behavior is really difficult to predict. Furthermore, to the best of these authors' knowledge, 
a good understanding of how porous medium micro-structure can enhance macroscopic transport is still lacking, especially for fibrous porous media. 
Many works on such media have been focused on the geometrical properties which can possibly affect standard Fickian dispersion and reaction, rather than on the intrinsic behavior of dispersion phenomena~\cite{torquato,tomadakis}.

In order to clarify this issue, this study presents results of 
several simulations at different preferential orientation
of fibers, porosity and Reynolds number. A
Lattice-Boltzmann-based model coupled with a Lagrangian particle tracking algorithm has been used.
The aim of the present paper is to clarify how the nematic properties of the porous medium
affects the mass and momentum transport mechanisms in order to design optimal porous
media with low drag and high effective mass diffusion. The minimization of drag reduces
the pump power demand, while the maximization of the mixing improves the homogeneity
of reacting species all along the porous medium, both effects enhancing the performance of
flow batteries. It will be shown, that porous media constituted by fibers preferentially oriented
along the flow direction exhibit smaller drag and higher effective diffusion.


\section{\label{sec2} Numerical Methodology}

During recent years the Lattice-Boltzmann Method (LBM) has gained much attention as alternative solution to the Navier-Stokes equations~\cite{succi}. Due to its numerical efficiency, easy parallelization and capability of handling complex geometries, the LBM is a promising tool to simulate complex flow fields at low Reynolds number, such as flows through fibrous electrodes for flow batteries. In the present study a  three-dimensional D3Q19 Lattice-Boltzmann  Multi-Relaxation Time (MRT) model has been implemented~\cite{mrt}. The MRT scheme allows to overcome some drawbacks of the Bhatnagar-Gross-Krook (BGK) formulation, which is the simplest and most common Lattice-Boltzmann equation, such as the viscosity-dependent numerical errors, especially in the case of very complex geometries~\cite{ginzburg}.
In order to simulate a pressure gradient $\Delta P / L$ in the flow, an equivalent body force  has been implemented. The Lattice-Boltzmann MRT equation reads as follows:
\begin{eqnarray}
f_r (\vec{x} + c_r \delta t, t+\delta t) - f_r(\vec{x},t) =  - \mathcal{M}^{-1} \big \{ \mathcal{S} \big ( m_r(\vec{x},t)-m^{eq}_r (\vec{x},t)   \big ) -  \big ( \mathcal{I} - \frac{1}{2}\mathcal{S} \big ) \big (\mathcal{M} F_r \big ) \big \} 
\label{lbmmrt}
\end{eqnarray}
where $f_r(\vec{x},t)$ is the distribution function along the $r$-th lattice direction at the position $\vec{x}$ and time $t$, $c_r$ is the so-called discrete velocity along the $r$-th direction, $\mathcal{M}$ is the transformation matrix, $\mathcal{S}$ the collision matrix, $\mathcal{I}$ the identity matrix, and $m_r$, $m_r^{eq}$ are the  moment and the equilibrium moment along the $r$-th lattice direction, respectively. Eq.(\ref{lbmmrt}) is a discrete formulation of the Boltzmann transport equation and states the relation between the collision step (right-hand of the equation) and the streaming step (left-hand of the equation). The set of moments $m_r$ consists of the hydrodynamic moments, which are conserved during collision, e.g. mass and momentum, and the non-conserved moments. In order to recover the correct Navier-Stokes  equation and avoid discrete lattice effects, the body force $F_r$ has been added during the collision step as follows~\cite{guo}:
\begin{eqnarray}
F_r=\omega_r \left ( \dfrac{ c_{r,i}-u_i}{c_s^2} + \dfrac{c_{r,i}\, u_i}{c_s^4}\, c_{r,i} \right ) \, \bigg ( \dfrac {\Delta P}{L} \bigg )_i \ ,
\end{eqnarray}
where $\omega_r$ is the weight of the LBM scheme along the $r$-th lattice direction and $c_{r,i}$, $u_i$ and $(\Delta P /L )_i$ are the  Eulerian component of the discrete speed, velocity, and pressure gradient, along the directions $i=x,y,z$. The macroscopic density $\rho$ and velocity $u_i$ are accurately recovered from the distribution functions $f_r$:
\begin{eqnarray}
\rho &=& \sum_r f_r \\
\rho u_i &=& \sum c_{r,i} \, f_r + \dfrac{1}{2} \dfrac{\Delta P}{L}\ .
\end{eqnarray}

 \begin{figure}
 \includegraphics[width=0.95\linewidth]{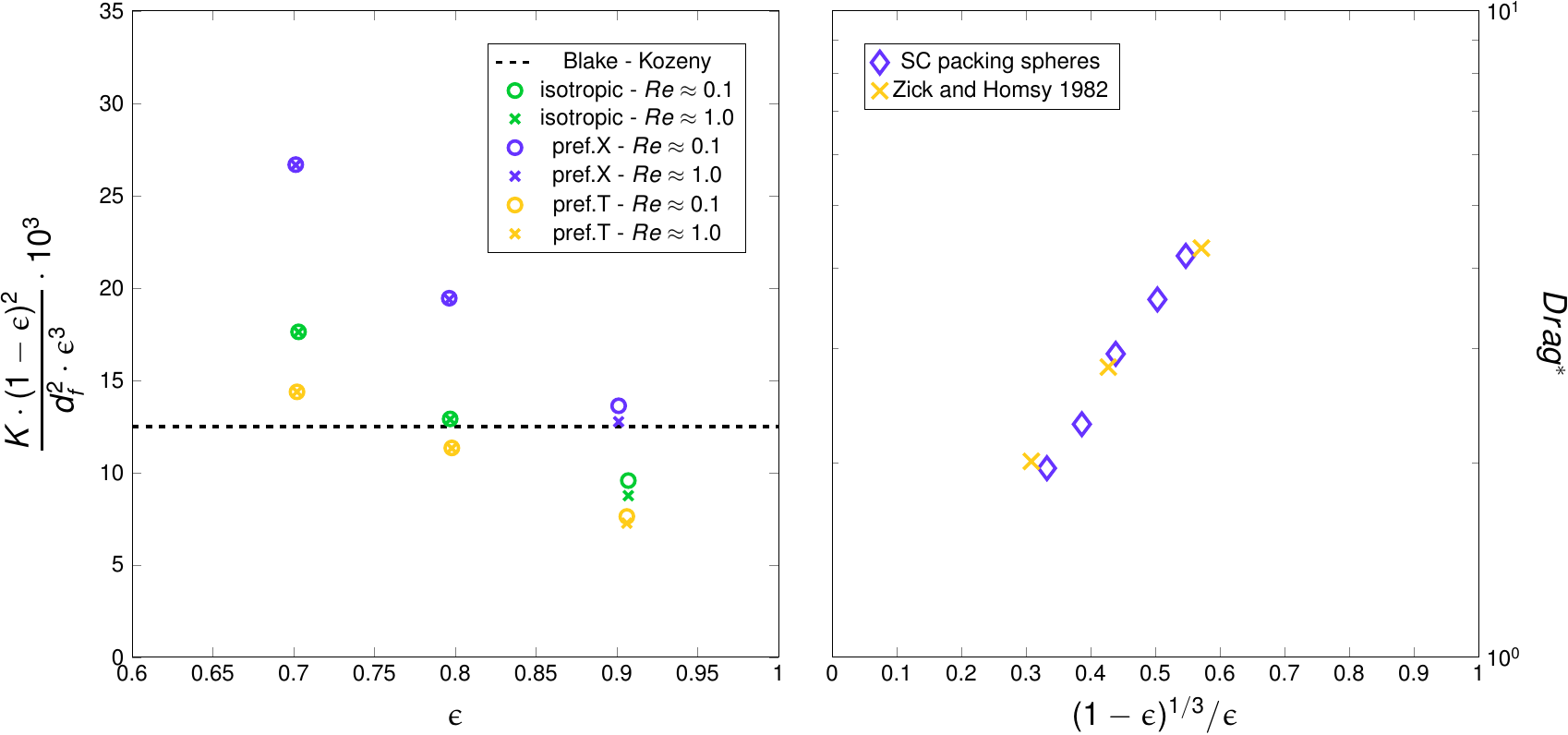}%
 \caption{\label{fig1} 
{Left Panel}: dimensionless permeability values for different cases of porosity $\epsilon$ and orientation. Results gather around the predicted value of the Blake-Kozeny equation, as expected. It should be pointed out that the permeability values diminish as the porosity increases, a trend already observed in Whitaker~\cite{whitaker-drag}.  {Right Panel: drag coefficients  $Drag^* =
2r_s^2/(9K) \, 1/(1-\epsilon)$ from numerical simulations of flows through single packed-bed of spheres of radius $r_s$ compared with results of Zick and Homsby}~\cite{zick}.}%
 \end{figure}

The present model is a further development of the Lattice-Boltzmann model already validated and used in Maggiolo et al.~\cite{maggiolo}. The model has been further validated by evaluating the permeability values obtained with different values of porosity, fiber orientation and Reynolds number {and the drag exerted by the flow on single packed-beds of spheres}. The permeability values $K$ have been obtained by means of the Darcy equation which relates the velocity with the pressure gradient:
\begin{eqnarray}
{K} &=& U \epsilon \, \mu \, \bigg ( \dfrac{\Delta P}{L} \bigg )^{-1} \ ,
\label{darcy}
\end{eqnarray}
where the pressure gradient $\Delta P/L$ corresponds to the applied body force, $U = [1/(\epsilon V)]\, \int_V u_x dV $ is the bulk intrinsic velocity of the generic cubic domain $V$ along the streamwise direction $x$, $\mu$ is the dynamic viscosity, and $\epsilon$ is the porosity.
Figure \ref{fig1} {shows values of dimensionless permeability against porosity on the left panel, and the values of drag coefficients of the packed bed  of spheres compared with the values of Zick and Homsy}~\cite{zick} on the right panel. Results of these simulations are considered proof of the  validity of the model. Dimensionless permeability values $K^*=K/d_f^2$ cluster around the permeability value of the Blake-Kozeny equation~\cite{whitaker-drag}:
\begin{equation}
{K^*} \, \dfrac{(1-\epsilon)^2}{ \epsilon^3} = \dfrac{1}{80} \ .
\label{permk}
\end{equation}

Eq.~\eqref{permk} has been derived for a medium composed of cylindrical fibers, for which  the equivalent diameter (i.e.\ six times the ratio between the volume of the fiber and its surface) equals three halves of the fiber diameter $d_f$~\cite{whitaker-disp}. Fig.~\ref{fig1} also shows that the higher the porosity, the lower the dimensionless permeability and the resulting pressure drop. 
The fiber Reynolds number has been determined as $\textit{Re}_f= U d_f / \nu$, where $\nu$ is the kinematic viscosity. Preliminary results of dispersion in these media show that the main dispersion features {do not depend on} the fiber Reynolds number, in the range $\textit{Re}_f=0.1\div1.0$, and on the porosity, in the range $\epsilon=0.7\div0.9$, {see the Appendix}~\ref{app}. Conversely, results instead show that dispersion depends on the orientation of the fibers. In light of this, further simulations have been carried out in order to characterize the effective dispersion induced by the porous medium micro-structure with high porosity values; the fiber Reynolds number has been fixed in the order of $\textit{Re}_f=10^{-1}$ and the porosity at $\epsilon=0.9$, in order to minimize drag.

\begin{figure}
 \includegraphics[width=1.0\linewidth]{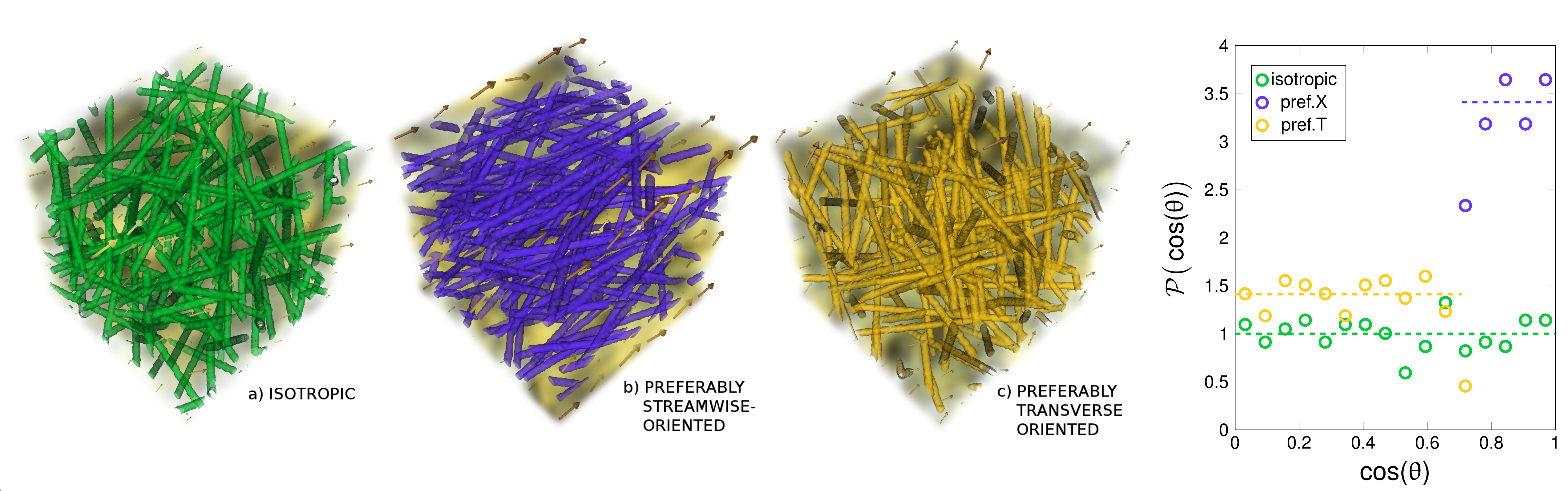}%
 \caption{\label{fig2}
{Left Panels}: Triperiodic cubic domains generated by numerical modeling where particles are randomly injected and advected by the steady flow field in the porous medium. Three main different orientations have been considered: a) an isotropic medium; b) a medium preferentially oriented along the streamwise direction $x$; c) a medium preferably oriented along the transverse directions $y$ and $z$. The medium is considered preferentially oriented along the streamwise or transverse direction when all the angles $\theta$ formed by the axes of the fibers and the $x$ axis are lower or higher than $\pi/4$, respectively. {Right Panel: probability distribution functions of $\cos(\theta)$ for different fibrous media.} 
}%
 \end{figure}

A triperiodic box has been considered, see Fig.~\ref{fig2}. The porous medium is composed of 50 fibers with length $\ell_f\approx160\sqrt{3}$ computational cells, uniformly random distributed throughout the domain. The side of the periodic box is discretized by $L=160$ Lattice-Boltzmann cells, while the fiber diameter $d_f$ corresponds to 6 cells. Three reference cases of fiber orientation have been investigated: an isotropic medium, an anisotropic medium preferentially oriented along the streamwise direction $x$ and an anisotropic medium preferentially oriented along the transverse directions $y$ and $z$. 
Seven isotropic, nine preferentially streamwise-oriented and eight preferentially transverse-oriented media have been considered in order to get a statistically significant samples, making a total of 24 simulations. The flow is driven along the $x$ direction by a mean pressure gradient $(\Delta P / L)$ which implies a fiber Reynolds number in the order $\textit{Re}_f=10^{-1}$. 

The mass transport properties have been evaluated following non-Brownian tracer particles injected in the fluid. The LBM stationary flow field has been used as input of the Lagrangian Particle tracking algorithm, and Lagrangian statistics of five thousand particles for each medium have been analyzed in order to extract the effective diffusion coefficients in the porous media. Particles have been randomly injected in the fluid phase and their trajectories have been stored in time. All the observables shown in the following analysis have been normalized using the bulk intrinsic velocity $U$  and the fiber diameter $d_f$.
 
In the whole text, the statistical observables (e.g.\ permeability $K^*$, effective viscosity ${\nu_{eff}}^*$,  Mean Square Displacement $MSD^*$) are calculated as mean values among all the cases belonging to one of the three main cases of fiber orientation, which are statistically equivalent, i.e.\ either isotropic, or preferentially streamwise-oriented, or preferentially transverse-oriented. This approach allows to determine the statistical confidence on the data.


\section{\label{sec3} Results}

\begin{figure}
 \includegraphics[width=0.7\linewidth]{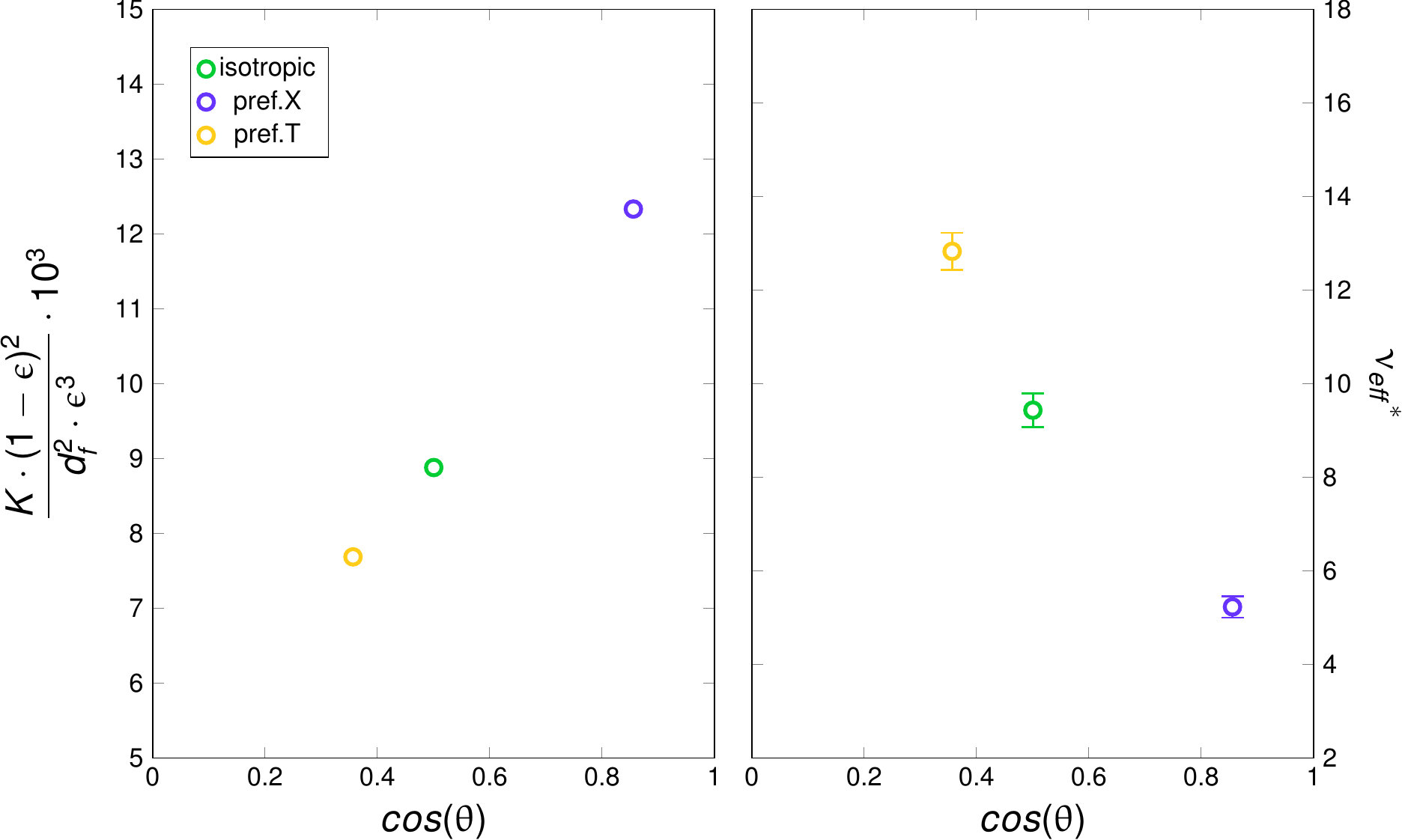}%
 \caption{\label{fig3}
Dimensionless permeability (left panel) and effective viscosity (i.e.\ ${\nu_{eff}}^*$, right panel), for the three main fiber orientations, plotted against the mean value of $cos(\theta)$. The effective viscosity is considerably reduced in the media preferentially oriented along the streamwise direction, indicating a significant reduction of the drag exerted on the medium, and, in turn, of the pressure drop. 
}%
\end{figure}

The bulk resistance to the flow induced by fibrous media with different typical fiber orientation is characterized by the permeability $K^*$ described in eq.~\eqref{permk}. At fixed flow rate, a higher value of permeability corresponds to a lower pressure drop, which is a desirable condition in RFBs {because it allows to reduce} the pump power needed to flow the electrolyte through the electrodes. The 
left panel of Fig.~\ref{fig3} shows the values of $K^*$ obtained in the three cases differing for the typical fiber orientation that are parametrized by the  mean value of $\cos(\theta)$, with $\theta$ {the angles between the fibers} and the $x$ axis.
The medium constituted by preferentially oriented fibers along the streamwise direction presents the highest value of permeability, 
while that with fibers preferentially transverse to the flow shows the lowest permeability. This behavior is not surprising considering that the former case presents the lowest projected area of the fibers on the cross-stream plane, while the latter the highest. 
In particular, it should be noted that media with fibers preferentially aligned along the flow reduce the overall drag by around 35\% with respect to isotropic fibrous media.

For later convenience, the same behavior is quantified in terms of an effective diffusion coefficient of the momentum transport. An effective (kinematic) viscosity $\nu_{eff}$ has been defined considering the ratio between driving force $({\Delta P}/{L})(1/\rho)$ and an effective viscous force given by $U/d_f^2$. In dimensionless form, the effective viscosity ${\nu_{eff}}^*=\nu_{eff}/(U d_f)$  reads:
\begin{eqnarray} 
{\nu_{eff}}^* = \bigg ( \dfrac{\Delta P}{L}\dfrac{1}{\rho} \bigg ) \dfrac{d_f}{U^2} \ .
\end{eqnarray}

Being the effective viscosity strictly related to the permeability, it conveys similar informations, however it will be useful to determine the dispersion efficiency, dimensionally homogeneous to the effective diffusion coefficient.
Although the general behaviors  of ${\nu_{eff}}^*$ and $1/K^*$  are equivalent, they are not proportional because the fiber Reynolds number $\textit{Re}_f$ slightly differs from the cases with constant driving force (Tab.~\ref{tab}).

\setlength\extrarowheight{5pt}
\begin{table}
\caption{\label{tab} Fiber Reynolds numbers and dimensionless root mean square velocities along the streamwise, $\langle {u_x'}^2 \rangle / U^2$, and transverse, $\langle {u_t'}^2 \rangle / U^2$, directions, for different fiber orientations.}
\begin{tabular}{| l | c | c | c |}
\hline
orientation & $\textit{Re}_f$ & $\langle {u_x'}^2 \rangle / U^2$ & $\langle {u_t'}^2 \rangle / U^2$ \\
\hline
isotropic & 0.12739 & 0.50348 & 0.10369 \\
pref.X & 0.17146 & 0.41988 & 0.06058\\
pref.T & 0.10911 & 0.49803 & 0.13800 \\
\hline
\end{tabular}
\end{table}

\begin{figure}
 \includegraphics[width=0.62\linewidth]{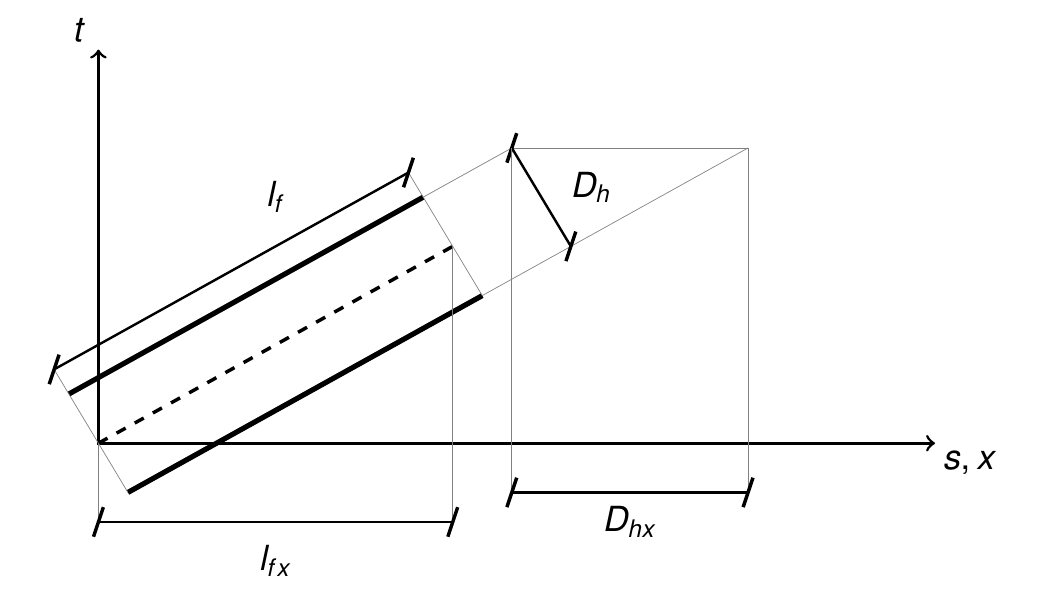}%
 \caption{\label{fig4} 
Hydraulic diameter 
$D_h$ and length $\ell_f$ of a ``tilted'' pore and their projections along the streamwise direction $D_{hx}$ and $\ell_{fx}$.
 } 
 \end{figure}

\begin{figure}
 \includegraphics[width=0.7\linewidth]{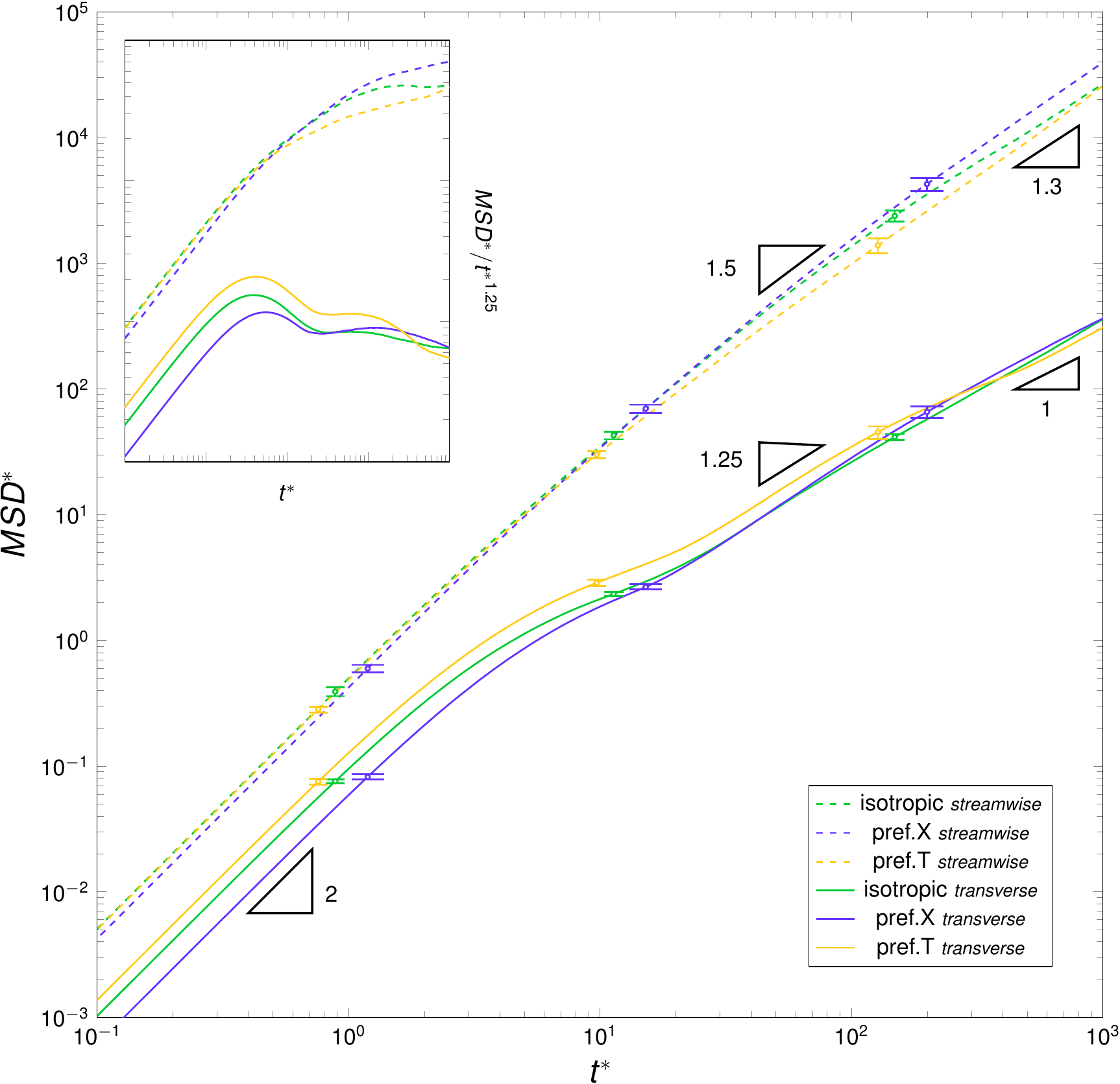}%
 \caption{\label{fig5} 
Dimensionless Mean Square Displacement against dimensionless time $t^*$ along the streamwise ($MSD_x^*$, dashed lines) and transverse ($MSD_t^*$, solid lines) directions. Three main regimes can be distinguished: 1) a ballistic dispersion for short times; 2) a superdiffusive dispersion for intermediate times; 3) a still superdiffusive dispersion along the streamwise direction and a Fickian standard dispersion along the transverse direction, i.e.\ $MSD^*_t\propto t$, for long times. The difference in the three main regimes is highlighted in the inset which shows the ratio ${MSD^*}/{t^*}^{1.25}$ approaching a constant value in the second and third regime along the transverse and streamwise direction, respectively.}%
 \end{figure}

Along with the minimization of the drag, in the case of flow batteries it is crucial to enhance the mixing of the electrolytes in the porous
medium constituting the electro-chemical active regions. To this purpose, it is important to characterize the dispersion properties of tracer particles distributed in the flow. The analysis has been restricted to non-Brownian particles whose trajectories coincide with those of fluid particles. As previously noted, 
the typical diffusion coefficient of electrolytes is quite small ($D\sim10^{-10}m^2/s^{-1}$) and usually negligible on the scale of the 
porous electrode. 
The main dispersion properties have been evaluated using the (dimensionless) Mean Square Displacements in the streamwise ($MSD^*_x$) and transverse ($MSD^*_t$) directions evaluated from the statistics of particle displacements $dx,dy,dz$ as follows:
\begin{eqnarray}
{{streamwise:}}\ \ &&MSD^*_x(t^*) = \langle {dx'}^2 \rangle / d_f^2  \\ 
{{transverse:}}\ \ &&MSD^*_t(t^*) = \big ( \langle {dy'}^2 \rangle + \langle {dz'}^2 \rangle \big )  / ( 2d_f^2) \ ,
\label{msd}
\end{eqnarray}
where $t^*=t\, U/d_f$ is the characteristic time, $dx(t^*)$ is the displacement along a generic direction $x$, 
$dx'(t^*)=dx-\langle dx \rangle$ is the displacement fluctuation, and $\langle \cdot \rangle$ indicates the ensemble averaging operator. 
Figure \ref{fig5} shows the values of the mean square
displacement $MSD^*$ along the streamwise and transverse directions as a function of the characteristic time $t^*$. The $MSD^*$ is 
initially proportional to ${t^*}^2$, corresponding  to a straight line with slope 2 in the log-log plot. This is expected  for small 
characteristic times when the particle motions are strongly correlated with ballistic trajectories. After a characteristic time $t^*
\approx10$, the behavior of the $MSD^*$ changes as $MSD^* \propto {t^*}^\alpha$, with $\alpha\approx1.50$ and $\alpha
\approx1.25$ along the streamwise and transverse directions, respectively. These exponents correspond to a superdiffusive 
dispersion behavior.   The superdiffusive dispersion process denotes a fast anomalous dispersion which has been found in biological and other specific physical systems~\cite{wang}. 
For longer characteristic times $t^*\gtrsim 100$ the dispersion behavior changes again.
The $MSD^*_x$ becomes proportional to $\approx {t^*}^{1.3}$ along the streamwise direction, i.e.\ the dispersion is still superdiffusive but with a different exponent, while it approaches a standard Fickian dispersion with the $MSD^*_t$  proportional to the time $t^*$  along the transverse direction. 
These different behaviors identifies three main regimes: (i) a typical ballistic dispersion for short times, (ii) a superdiffusive dispersion for intermediate times, and (iii) a third regime for long times which is Fickian {in the transverse direction}, while still superdiffusive for the streamwise direction.
The hydraulic diameter $D_h=4\epsilon V/S\approx9\,d_f$ has been related to the mean intercept length in porous media~\cite{dullien}.
More to the point, it indicates the mean width of the pores formed by the porous micro-structure. In order to take into account the effect of the fibers orientation, one should consider the main pores width and length along the streamwise $x$ direction, that is, ${D_h}_x=D_h/\sqrt{1-cos(\theta)^2}$ and ${l_f}_x=l_f\ cos(\theta)$, {see Fig}~\ref{fig4}. Following this approach the mean half-width of the pores (i.e.\ the main radius of the pores) and their mean length result ${R_h}_x=4.8d_f, \ 5.2d_f, \ 8.8d_f$ and ${l_f}_x=15d_f, \ 23d_f, \ 40d_f$ for the transverse oriented, isotropic and streamwise oriented medium, respectively.
In the light of this, the three different behaviors can be interpreted with typical time scales,  estimated considering the bulk velocity and the porous radius $\approx {R_h}_x$ and the porous length $\approx {l_f}_x$ induced by the aspect ratio of the fibers. 
When $t^*<{R_h}_x/d_f$ particles are traveling inside a pore width, possibly rounding one fiber, while, when $t^*\gg {l_f}_x/d_f$ particles have crossed the whole pore length and their dispersion becomes Fickian, on the average. Conversely, when ${R_h}_x/d_f < t^* \lessapprox {l_f}_x/d_f$, particles are traveling along the anisotropic pores and their trajectories tend to follow the pores axes, resulting in an enhancement of the dispersion in  the medium.

It should be noted  that different exponents of the time behavior in $MSD^*$ have been reported in the literature, considering different porous media, see e.g.~\cite{kang2015}.
The present data, showing long-time superdiffusive streamwise dispersion, are consistent with the recent findings of Kang et al.~\cite{kang} who analyzed a numerical simulation of a real porous geometry reconstructed by a micro-tomography. 
Concerning the difference among the three cases considered which differ for the fiber orientation, a very similar 
trend for the dispersion in the streamwise direction for all cases has been noted. 
However for the long time behavior, the transition from the superdiffusive dispersion with $\alpha\approx1.5$ to that with  $\alpha\approx1.3$ occurs at shorter times for the isotropic and the preferentially transverse cases so the particles of the preferentially streamwise aligned medium show eventually a slightly larger $MSD^*_x$ for high $t^*$. 

Concerning the differences found in the transverse dispersion, in the ballistic short-time behavior, the transverse medium shows higher rate of dispersion, followed by the isotropic case and {then the streamwise} oriented one. In this limit, $MSD^*_t\propto \langle{u_t'}^2\rangle/U^2\, {t^*}^2$, so that ${t^*}^2$ ballistic dispersion is imposed by the fluid cross-stream velocity fluctuation levels which {are maximal} in the transverse case (Tab.~\ref{tab}). 
After the ballistic motion, all the cases show a similar time scale with transition to the super-diffusive behavior. 
However, the streamwise oriented medium shows a slightly longer transition time. 
A similar trend is shown for the second transition from the super-diffusive to the regular Fickian behavior, where again the streamwise oriented medium shows the longest transition time. 
The longer permanence time of this case in the ballistic and  super-diffusive behaviors allows a recovery of the slower dispersion shown for small $t^*$. 
For this reason, all cases show a similar level of $MSD^*_t$ for long time, when they exhibit a Fickian diffusion in the transverse direction. The porous structures created by the streamwise oriented direction, which is possibly able to correlate the motion of a fraction of particles for a much longer time, provides a possible explanation.

To this purpose, it is important to analyze the correlations of the particle motion in order to understand the different behaviors. The dimensionless autocorrelation functions ${c'_v}^*$ along the streamwise and transverse directions have been determined as follows:
\begin{eqnarray}
{{streamwise:}}\ \ &&{c'_v}_x^*(t^*) =  \dfrac{\langle v'_x(t^*)\, v'_x(0) \rangle}{\langle {v'_x(0)}^2 \rangle} \\ 
{{transverse:}}\ \ && {c'_v}_t^*(t^*)  = \dfrac{1}{2} \bigg (  \dfrac{ \langle  v'_y(t^*) \, v'_y(0) \rangle}{\langle {v'_y(0)}^2 \rangle} + \dfrac{\langle v'_z(t^*) \, v'_z(0)  \rangle}{\langle {v'_z(0)}^2 \rangle} \bigg )
\label{vcorr}
\end{eqnarray}
where $v_{i}$ and $v'_{i}=v_{i}-\langle v_{i} \rangle$ are the particle absolute and fluctuation velocities along the direction $i=x,y,z$.

\begin{figure}
 \includegraphics[width=0.7\linewidth]{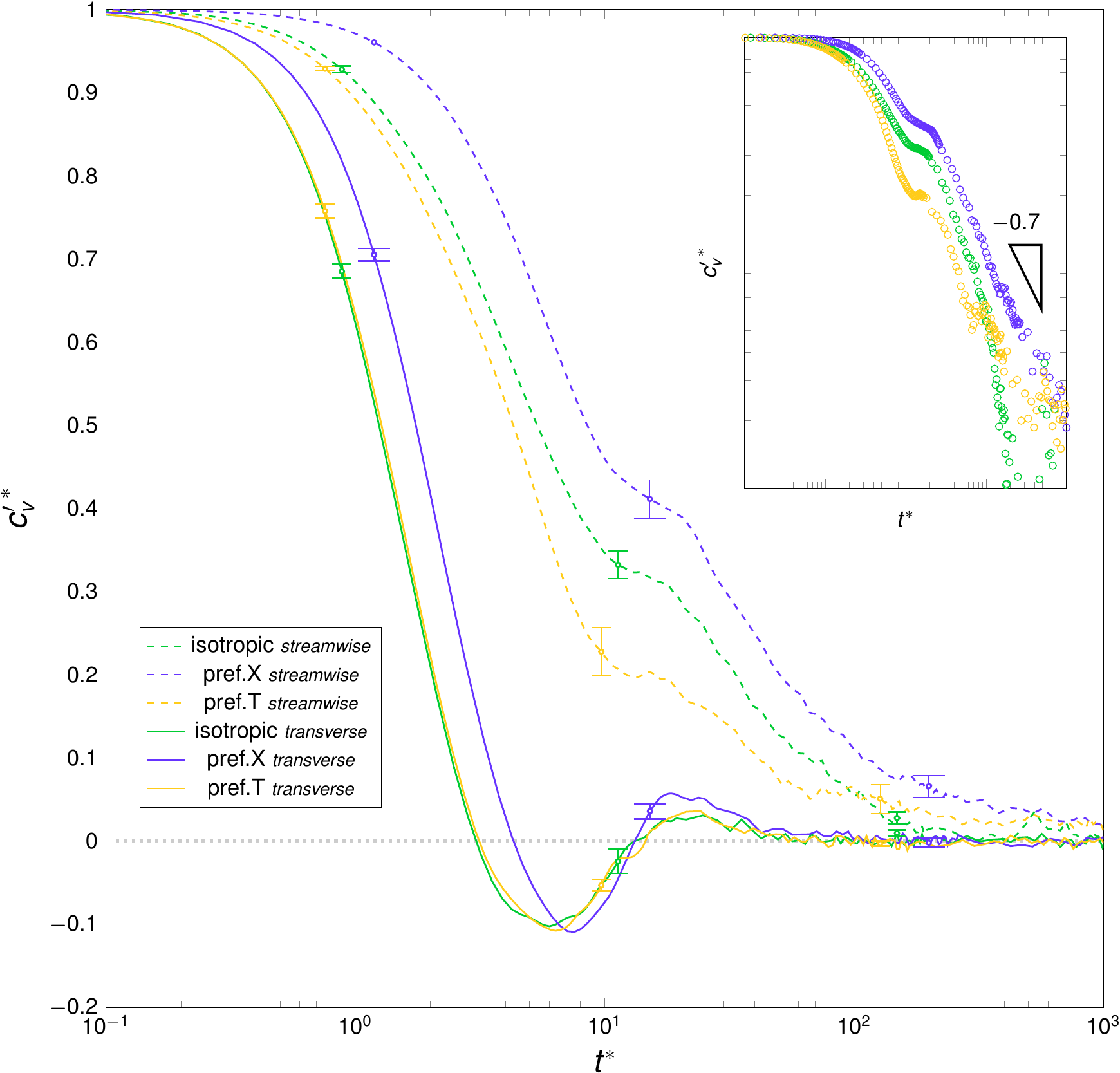}%
 \caption{\label{fig6}
Dimensionless velocity autocorrelation ${{c'_v}^*}$ against dimensionless time $t^*$ along the streamwise (${{c'_v}_x^*}$, dashed lines) and transverse (${c'_v}_t^*$, solid lines) directions. The autocorrelation approaches zero along the transverse direction for long characteristic times, i.e.\ in the third regime observed. The figure highlights that the streamwise-oriented medium presents higher velocity autocorrelations not only along the streamwise direction (as expected since fibers are aligned with the main flow), but also in the transverse direction. The inset represents velocity autocorrelations along the streamwise direction in log-log plot: it can be observed that ${c'_v}^*_x$ decreases proportionally to ${t^*}^{-0.7}$ towards the end, which in turn corresponds to an increasing of the $MSD^*_x \propto {t^*}^{1.3}$.
}%
 \end{figure}

Results of dimensionless autocorrelation functions ${c'_v}^*$ for different fiber orientations are plotted against the characteristic time $t^*$ in Fig.~\ref{fig6}. The velocity autocorrelation ${c'_v}_x^*$ along the streamwise direction is significantly higher than the transverse one for all the considered cases. Moreover, for long times, the streamwise velocity autocorrelation slowly decays, while that for the transverse direction decorrelates for $t^*\approx 100 \gg {l_f}_x/d_f$. This behavior reflects the anisotropy induced by the mean flow driven by the mean pressure gradient. 
The correlation of the streamwise motion for longer times is attributed to two typical particle trajectories,
the former to the particle that travels near a stagnation point, the latter to particles which flow in the bulk of some almost streamwise aligned long pores. In both cases the particles tend to reside for long time in the same state characterized by a similar
velocity. 
Before discussing the effect of the preferential fiber alignment, it is interesting to note how different is the autocorrelation of
the transverse velocity. It shows a faster decorrelation and shows a negative minimum around $t^*\sim5\div9$, which approximately corresponds to the mean half-width of pores in the medium ${R_h}_x$. 
This minimum indicates that after that time, the particle motion tend to reverse in the transverse direction and this can be interpreted as an effect induced by the particles which are rounding the fibers. 
Actually after the minimum, the autocorrelation shows a relative maximum before decorrelating, which can be interpreted as the effect of a successive fiber encountered during the motion. 
Concerning the effect of the preferential fiber orientation, the autocorrelation of the preferentially streamwise oriented fibers is in general higher for longer times for both the streamwise and transversal motions. In some sense, it appears that the pores created by the almost streamwise fiber are able to be followed by a part of particles for longer times. 
This induces {a higher correlation} for both velocities, since the pores are not completely streamwise oriented along the weakly tilted fibers. This results in longer characteristic times which divides the typical dispersion behaviors observed in the $MSD^*$.
As it is {well known}, the mean square displacement is actually determined by the autocorrelation function as,
\begin{equation}
MSD^*={2}\, \langle {v'(0)}^2\rangle\, t^*  \int_0^{t^*}  \bigg (1-\dfrac{s}{t^*} \bigg ) {c'_v}^*(s)\ d{s} \ .
\label{msdac}
\end{equation}

From eq.(\ref{msdac}), the ballistic behavior $MSD^*{\propto t^*}^2$ can be obtained for  ${c'_v}^*=1$, while for high $t^*$ and 
${c'_v}^*=0$ the dispersion becomes regular and Brownian, namely $MSD^*{\propto t^*}$. If instead for long times it is
${c'_v}^*\propto {t^*}^{\beta}$, the mean square displacement shows an anomalous diffusion, $MSD^*{\propto t^*}^\alpha$ with 
$\alpha=2+\beta$. Hence, the different dynamics noted in the $MSD^*$ analysis reflects in different behaviors of the autocorrelation. 
It has been previously noticed how the preferentially streamwise oriented medium shows longer characteristic times for the transition 
among the different dynamics and that this reflects in higher dispersion at longer times. Moreover, the long time behavior
observed for the mean square displacement is consistent with the results of the autocorrelation where a decorrelation is found
for the transverse velocity and a slow decay with $\beta\simeq=-0.7$ is observed for the streamwise direction, as shown in Fig.~\ref{fig6}.

\begin{figure}
 \includegraphics[width=0.8\linewidth]{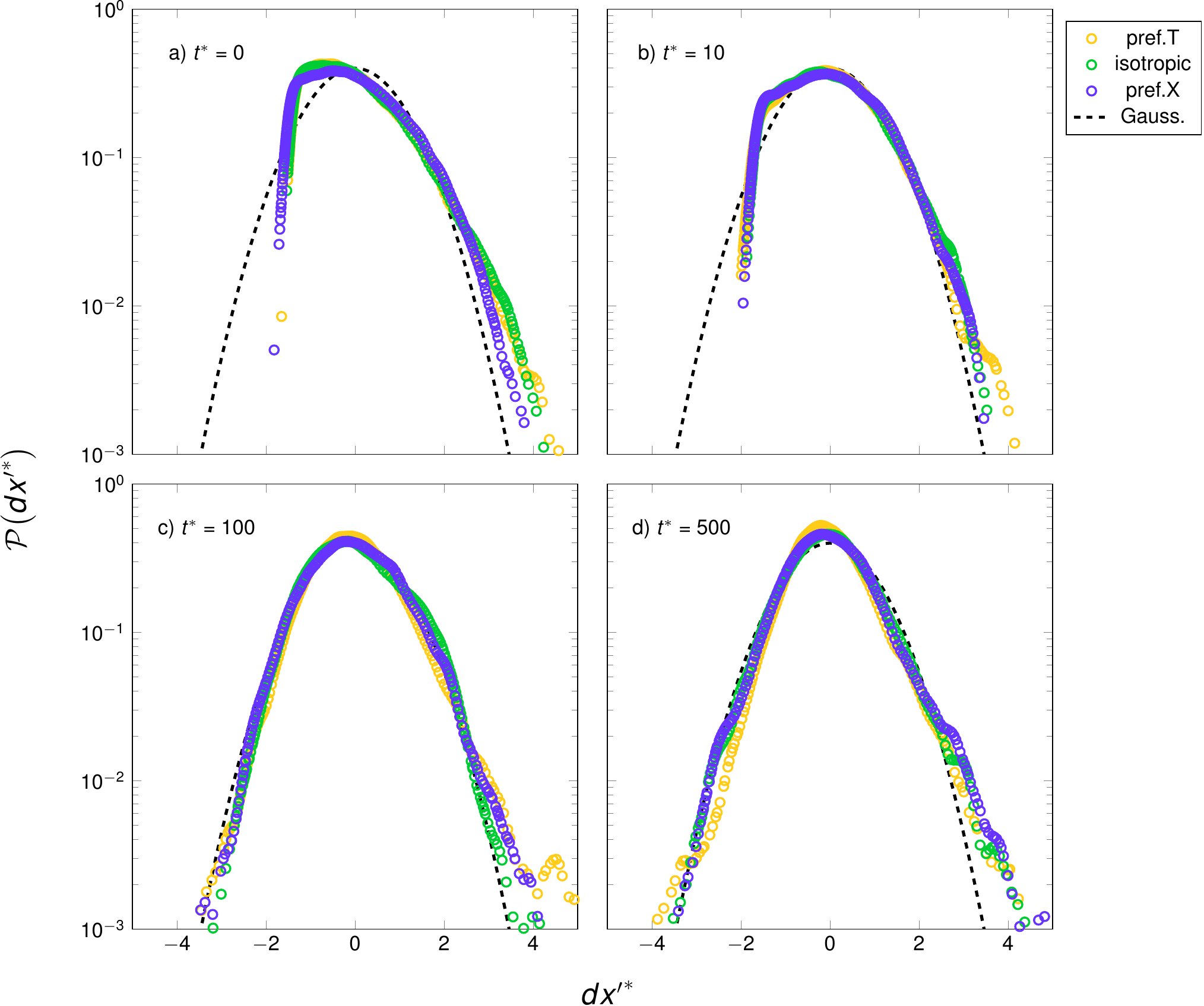}%
 \caption{\label{fig7}
Probability distribution functions (PDF) of dimensionless displacement fluctuations, $ d{x'}^* = ( dx- \langle dx \rangle ) / \sigma (dx)$ (where $\sigma(dx)$ is the variance), along the streamwise direction, for different fiber orientations. (a) PDF at $t^*=0$; (b) PDF at $t^*=10$ ; (c) PDF at $t^*=100$; (d) PDF at $t^*=500$. 
}%
\end{figure}

\begin{figure}
 \includegraphics[width=0.8\linewidth]{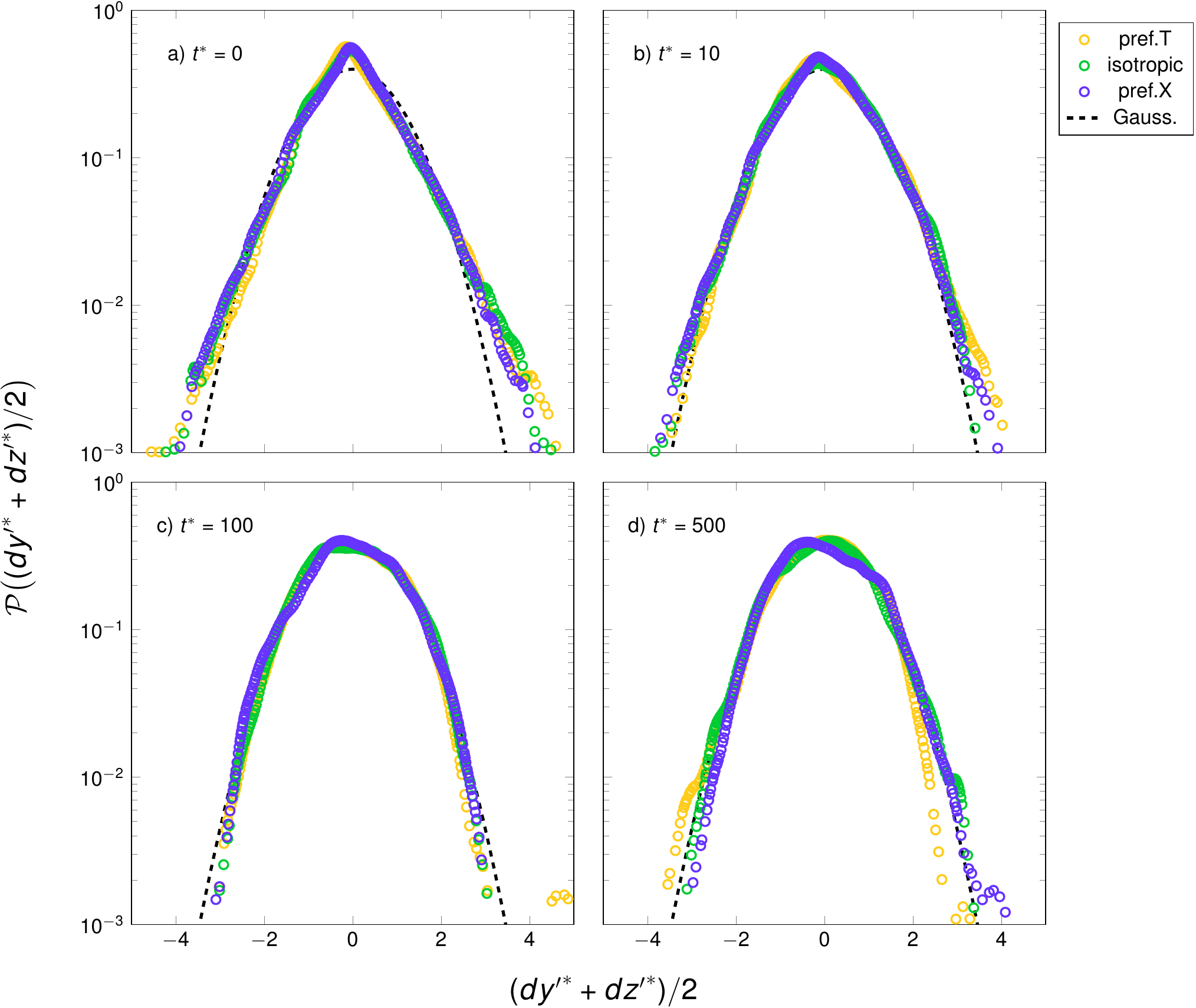}%
 \caption{\label{fig8}
Probability distribution functions of dimensionless displacement fluctuations, $ (d{y'}^*+d{z'}^*)/2 = \big( ( dy- \langle dy \rangle ) / \sigma (dy) + ( dz- \langle dz \rangle ) / \sigma (dz) \big ) /2$  (where $\sigma$ is the variance), along the transverse direction, for different fiber orientations. (a) PDF at $t^*=0$; (b) PDF at $t^*=10$ ; (c) PDF at $t^*=100$; (d) PDF at $t^*=500$. 
}%
\end{figure}

To better highlight the origin of the different anomalous  dispersion behaviors, the
 Probability Distribution Functions (PDF) of the particle displacements along the streamwise and transverse directions 
 are shown in Figures \ref{fig7} and \ref{fig8}. {The displacements are relative to the mean position and normalized 
 with the square root of the variance
 ($MSD^*$) in order to highlight the difference with the corresponding Gaussian distribution}.
At $t^*=0$ and for small $t^*$ the PDF of the displacements correspond to that of the flow field $u_x$, since the particle motion is highly correlated with their initial conditions. 
Focusing on the streamwise behavior first, see Fig.~\ref{fig7} panel a), the displacements and the underlying velocity field is 
strongly non-Gaussian. The mean flow direction induces a highly asymmetric PDF with a steep positive tail. In particular the 
streamwise velocity field is characterized by low-velocity regions (near stagnation points) and by high-velocity preferential paths which correspond to long pores mainly oriented along the streamwise direction. This behavior induces a great asymmetry of the displacement at small times with a PDF which is left-truncated and reveals a significant part of particles traveling at speed far higher than the average (right tail of the PDF). Since at $t^*\approx10$ the streamwise motion is still  correlated (Fig.~\ref{fig6}), the PDF is very
similar to the initial one.  
After a characteristic time $t^*\approx100$ , which identifies the end of the ballistic motion with a power-law decaying correlation, the shape of the PDF starts to approach the Gaussian shape, see panels $c$ and $d$ of Fig.~\ref{fig7}, with still a positive tail slightly 
steeper than a Gaussian. No significant differences emerge from PDFs of different media. 
The origin of the streamwise anomalous diffusion appears related to the skewed non-Gaussian intermittent flow velocity field induced by the fibrous microstructure. 
However, the different fiber preferential alignment appears to alter the typical correlation time scales of the flow and not the single-point statistical behavior.

The PDF of transverse particle displacements is  represented in Fig.~\ref{fig8}. The transverse flow velocity field PDF corresponds to the PDF at $t^*=0$,
panel a), and appears non-Gaussian, but symmetric because of the isotropy of motions in the cross-stream plane. 
The fibrous microstructure induces very steep tails in the PDF at $t^*=0$ denoting that the pores are able to transport
particles at longer transversal distance with respect to a corresponding Gaussian process. 
At $t^*\simeq10$, being the motion in the transversal direction uncorrelated, the displacements PDF differs from the initial one and
in particular results more Gaussian. At $t^*\simeq100$ and $t^*\simeq500$ is even more Gaussian. 

\begin{figure}
 \includegraphics[width=0.72\linewidth]{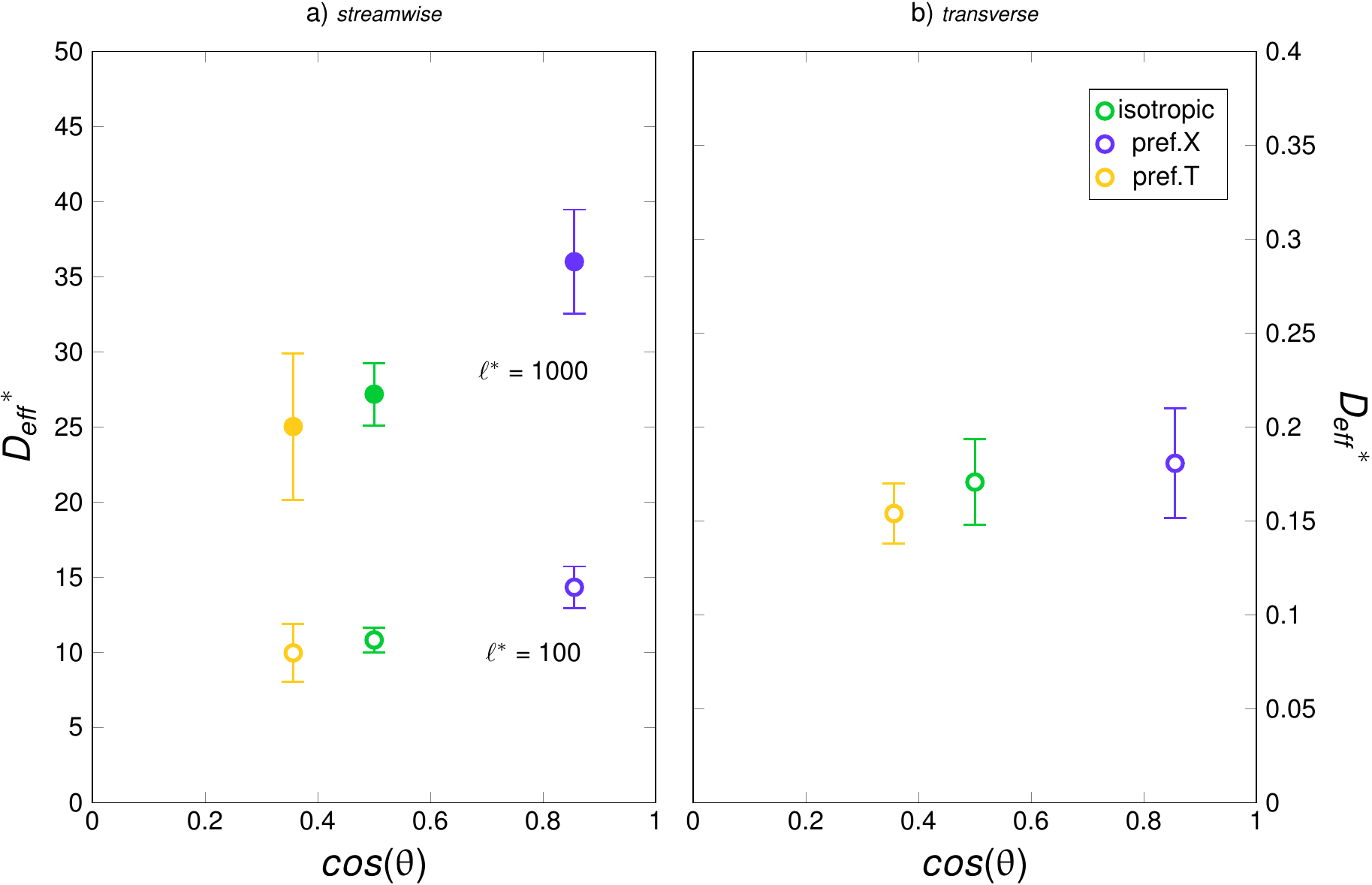}%
 \caption{\label{fig9}
Dimensionless equivalent effective dispersion coefficients ${D_{eff}}^*$ plotted against the mean value of $cos(\theta)$, for different cases of fibers orientation, along the streamwise (left panel) and transverse (right panel) direction. In left panel void marks indicate a characteristic displacement length $\ell^*=\ell/d_f=100$, whereas filled ones indicate $\ell^*=1000$.
}%
 \end{figure}

Hence concluding, the dispersion dynamics in fibrous media with different preferential alignment of the fibers show 
a long-lasting superdiffusive behavior of the streamwise dispersion which appears induced by the highly non-Gaussian velocity 
field. 
On the contrary, in the transverse directions, a long-time regular diffusion has been observed. 
On the shorter time-scales, the dispersions induced by fibrous media are superdiffusive for both directions. 
In these regime we found the most important difference among the cases differing for the
fiber preferential alignment. The typical time-scale at which these highly dispersive behaviors end is longer when the fibers are
preferentially streamwise oriented. This results in a overall mean square displacement for longer times ($MSD^*$) which is  
higher or similar to when the medium is constituted by isotropic oriented fibers. 
In other words, aligning the fibers preferentially along the streamwise direction slightly increases the long-time dispersion. 
To give an overall measure of the long time dispersion, the dimensionless dispersion coefficients ${D}^*$ have been computed by fitting the dimensionless mean square displacements $MSD^*=2{D}^*{t^*}^\alpha$.
When dealing with macroscopic transport of species inside porous media in applications, it is useful to model the mass transport via convection-diffusion-reaction Eulerian equations for the homogenized volume. When a regular diffusion process ($\alpha=1$) takes place, the only parameter needed is the mass diffusion coefficient ${D}^*$. Nonetheless, when $\alpha \neq 1$ the corresponding Eulerian transport equation consists of nontrivial fractional derivatives~\cite{schumer}. In order to overcome this issue at the practical aim, it is convenient to transform the superdiffusive dispersion process in an equivalent regular dispersion process so as to solve a usual convection-diffusion-reaction equation for the Eulerian homogenized mass transport. However, in order to define an equivalent Gaussian system {it is necessary} to fix a typical displacement length $\ell^* := \ell/d_f$ that characterizes the typical size of the system where the anomalous dispersion process occurs. The equivalent system is then obtained matching the actual Mean Square Displacements calculated at the system size, i.e.\ $MSD^* = {\ell^*}^2$, with the equivalent normal diffusion process characterized by ${D_{eff}}^*$.
The effective mass diffusivity has thus been derived by imposing the following equality:
\begin{equation}
{\ell^*}^2 = 2D^*{t^*}^\alpha=2{D_{eff}}^*t^* \ ,
\end{equation}
from which it follows:
\begin{equation}
{D_{eff}}^*(\ell^*) = 2^{\frac{1-\alpha}{\alpha}} {D^*}^{\frac{1}{\alpha}}{\ell^*}^{2\frac{\alpha-1}{\alpha}}
\end{equation}

Figure \ref{fig9} well depicts the effect of fibers orientation on the streamwise and transverse dispersion. Values of ${D_{eff}}^*$ are plotted against the mean value of $cos(\theta)$. While no significant differences on the dispersion coefficient are observed along the transverse direction, an increment of dispersion can be achieved along the streamwise direction by choosing to align fibers preferentially along the streamwise direction. Being the process superdiffusive, the larger the typical system scale, the higher the
effective diffusion coefficient. E.g.\ considering the typical streamwise extension of the system in the order of   1000 fiber diameters,
the effective diffusion coefficient becomes 100 times larger than that in the transversal direction. 

It should be remarked that the medium with fibers aligned along the flow exhibits the lowest resistance to the flow, i.e.\ the highest 
permeability. At practical purposes, it is crucial to define an efficiency in terms of ratio between energy spent to drive the flow and mixing achieved in the porous medium.  
This efficiency as a function of the preferential orientation of the fibers in the medium can be expressed as an effective Schmidt number $\textit{Sc}$ which corresponds to the ratio between the effective viscosity and  diffusion coefficients:
\begin{equation}
\textit{Sc} = \dfrac{ {\nu_{eff}}^* }{ {D_{eff}}^* } 
\ . \label{sch}
\end{equation}

\begin{figure}
\includegraphics[width=0.72\linewidth]{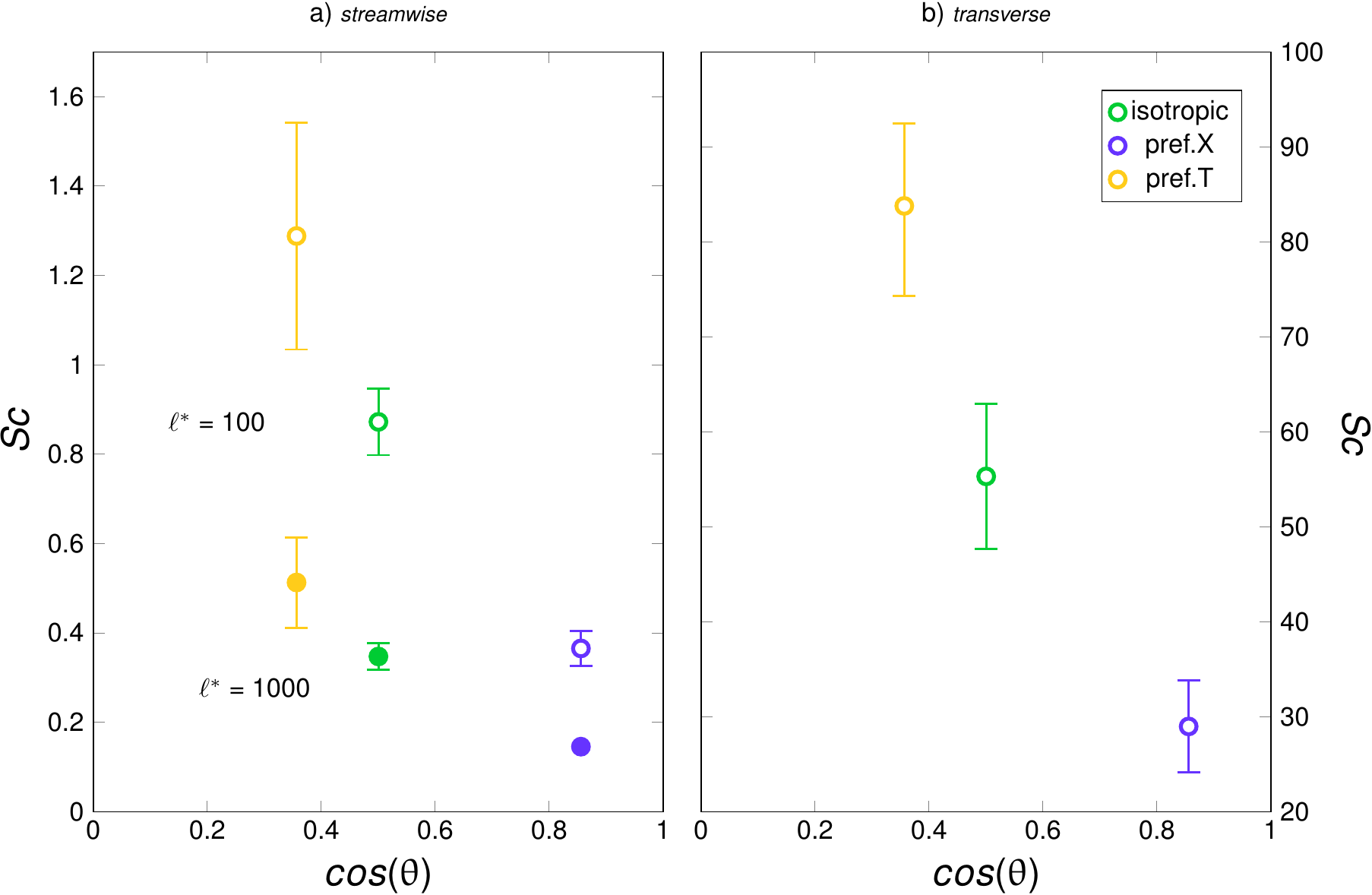}%
 \caption{\label{fig10}
Schmidt number $\textit{Sc}$ plotted against the mean value of $cos(\theta)$. Both along the streamwise and transverse directions the medium preferentially oriented along the streamwise direction $x$ presents the lower value of Schmidt number $Sc$ and, consequently, the highest dispersion efficiency.
}
 \end{figure}

The values of effective Schmidt numbers has been reported in Fig.~\ref{fig10} which highlights the dispersion efficiency of the differently oriented media. The lower the Schmidt number, the lower the ratio between the effective viscosity and the effective dispersion. In other words, low Schmidt numbers indicate low drag and high dispersion, which in turn increases the electrodes performances. The values of $\textit{Sc}$ have been evaluated for the same pressure gradient $\Delta P / L$ rather than for the same $\textit{Re}_f$, since the same pressure gradient imposes slightly different Reynolds numbers (Tab.~\ref{tab}). This is useful for practical applications since the main tunable parameter in real flow batteries systems is $\Delta P / L$. In other words, the effective Schmidt number indicates how much efficient is the dispersion in the fibrous medium, at a fixed $\Delta P / L$. 

\begin{figure}
\includegraphics[width=0.72\linewidth]{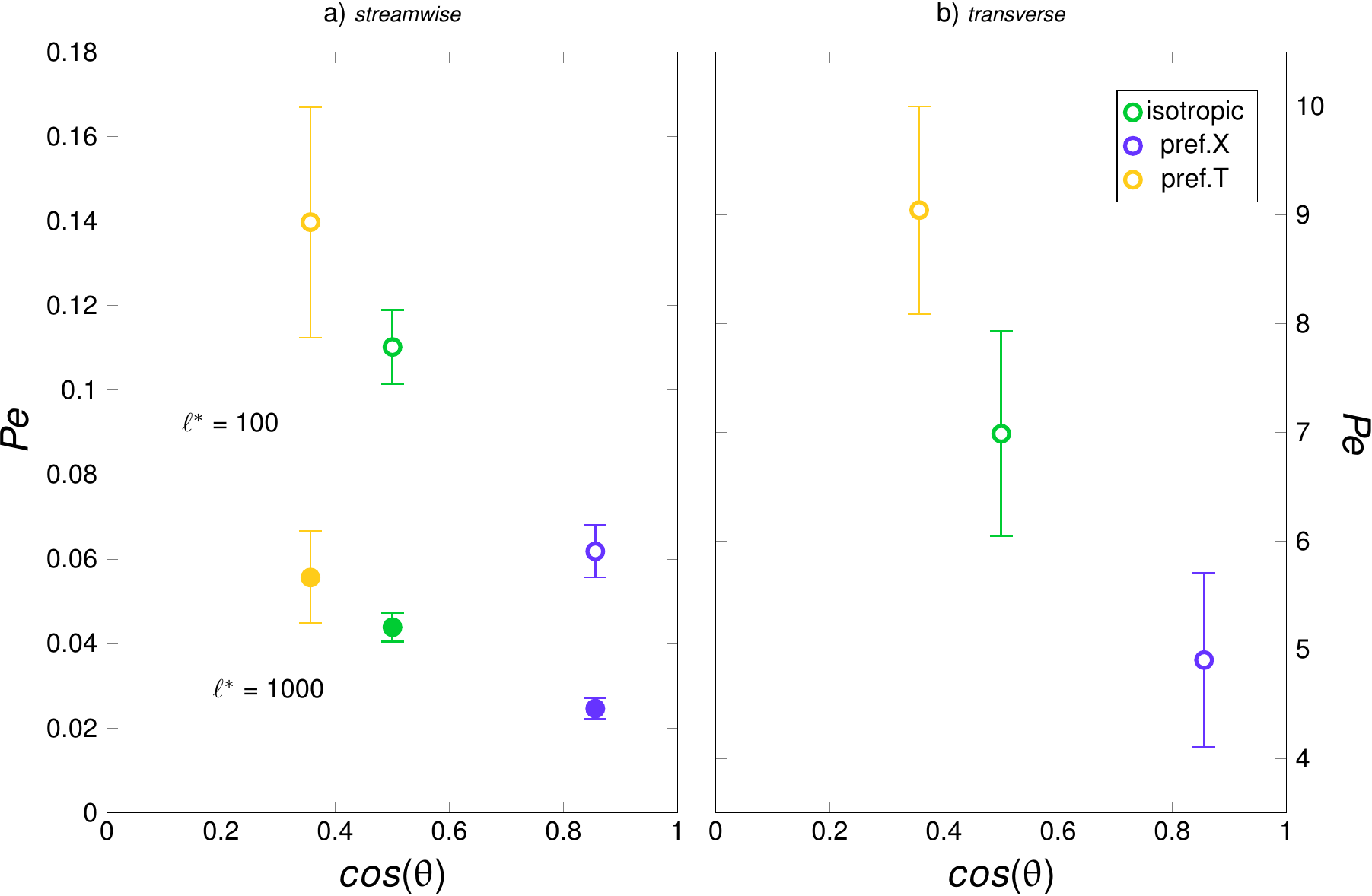}%
 \caption{\label{fig11}
 Peclet number $\textit{Pe}$ plotted against the mean value of $cos(\theta)$. Both along the streamwise and transverse directions the medium preferentially oriented along the streamwise direction $x$ presents the lowest value of Peclet number.
}
\end{figure}

It should be noted that the $\textit{Sc}$ number depends on the fiber Reynolds number since the dimensionless effective viscosity ${\nu_{eff}}^*$ is inversely proportional to it, while the dimensionless diffusion coefficient ${D_{eff}}^*$ is independent of it. 
From a theoretical point of view, it is interesting to note that the effective Peclet number $\textit{Pe}$ defined as, 
\begin{equation}
\textit{Pe} = \textit{Sc}\, \textit{Re}_f 
= \dfrac{\epsilon}{K^*} \dfrac{1}{{D_{eff}}^*} 
\ . \label{pec}
\end{equation}
gets rid of the dependence on the fiber Reynolds number since for viscous flow in porous media the dimensionless permeability is independent of the Reynolds number. 
Figure~\ref{fig11} depicts values of the effective Peclet number. This indicator shows an increase of dispersion efficiency less pronounced but still remarkable.  

The  behavior of the Schmidt and Peclet numbers unequivocally show that fibrous media preferentially oriented along the streamwise direction are characterized by higher effective mixing with the lowest pressure loss and appear the optimal configuration for porous electrodes for RFBs.

\section{\label{sec4} Final remarks}

In the present study, the effects of altering the micro-structure of a porous medium composed of fibers have been quantified in terms of permeability and effective dispersion features. 
In particular, the effect of the fiber orientation has been found to play a major pivotal role not only in the momentum transport behavior across the medium, but even in the dispersion dynamics. These results are crucial to optimize the fibrous electrodes of Redox-Flow-Batteries. 

Microscale modeling of viscous flows through differently oriented fibrous media have been performed  by means of a numerical algorithm based on the Lattice-Boltzmann method. Three main categories of fibrous media have been considered: isotropic and preferentially aligned and transversal with the bulk flow. The evolution of passive tracer particles dispersed in the flow has been obtained using a Lagrangian Particle Tracking algorithm. 

Results show that, as expected, the permeability value of the medium $K^*$ is increased by preferentially aligning the fibers along the streamwise direction so the overall drag exerted on the flow is diminished. The opposite behavior is observed for the media with fibers transverse to the flow. 

The  Mean Square Displacement $MSD^*$  of tracers show different anomalous and regular dispersion behaviors both along the streamwise and transverse directions. 
Three different regimes have been identified: (i) a ballistic dispersion for very short characteristic times, (ii) a superdiffusive dispersion for intermediate characteristic times and (iii) a still superdiffusive dispersion and a standard Fickian dispersion along the streamwise and transverse directions, respectively, for long characteristic times. 

The effect of fiber orientation on the dispersion dynamics has found to be less pronounced in comparison to that on momentum transport. Nevertheless, the medium with fibers preferentially oriented along the streamwise direction shows slightly higher dispersion along both directions. This behavior has been ascribed to the capability of the latter medium to create pores which could be followed by tracers for longer time in comparison with the isotropic case, since the observed characteristic times of the autocorrelation functions are longer.

Actually, for energy storage applications, such as flow-batteries, the optimal medium should presents both low drag and high dispersion features. Therefore, the values of the effective Schmidt and Peclet numbers which characterize the dispersion efficiency have been determined. The Schmidt numbers give the ratio between the effective diffusion and the overall drag, but depends on the fiber Reynolds number. The Peclet number is instead independent of it. Results clearly show the higher efficiency that can be achieved by aligning the fibers preferentially along the streamwise direction. In particular, by adopting this strategy, one can increase approximately the dispersion efficiency {of 60\% and 45\%} along the streamwise and transverse directions, respectively, in terms of Schmidt number, and {of 40\% and 25\%} along the streamwise and transverse directions, respectively, in terms of Peclet number, in comparison with an isotropic medium. These numbers can be seen as the fluid mechanics efficiency.

For the sake of completeness, it can be shown that the dimensional effective dispersion coefficients $D_{eff}\, [m/s]$, evaluated from the calculated dimensionless values ${D_{eff}}^*$, overwhelm the molecular diffusion coefficients when real electrodes are considered. 
With regards to All-Vanadium Redox Flow Batteries, the kinematic viscosity and the typical fiber diameter are $\nu_{v}=4.4\cdot 10^{-6} \, [m^2/s]$ and ${d_f}_v=10\div50\, [\mu m]$, respectively, whereas the value of typical displacement length $\ell$ is of the order of centimeters along the streamwise directions, so that $\ell^*\approx 1000$ ~\cite{tang}.
The dispersion coefficients can be thus evaluated as $D_{eff}={D_{eff}}^*(\ell^*)\, \textit{Re}_f \, \nu_{v}$.
Along the flow direction the dispersion coefficients are $D_{eff}=1.5\cdot 10^{-5}\, [m^2/s]$ and $D_{eff}=2.7 \cdot 10^{-5}\, [m^2/s]$, for the isotropic and the preferentially streamwise-oriented medium, whereas along the transverse direction they result $D_{eff}=0.96\cdot 10^{-7}\, [m^2/s]$ and $D_{eff}=1.36\cdot 10^{-7} \, [m^2/s]$, respectively. It should be noted that the typical molecular diffusion coefficient of Vanadium ions in water is of the order of
$D=10^{-9}\div10^{-10}\, [m^2/s]$ so several order of magnitude smaller than the effective diffusion coefficient promoted by the micro-structure of porous media. 
The results also show that the widely used Bruggeman correction~\cite{} to account for the effect of the porosity on the molecular diffusion,
$D_{b}=\epsilon^{3/2} D$, is negligible for the dispersion of electrolytes in water being $D_{b}\ll D_{eff}$ \footnote{Different is the case when gases in porous media are considered. The typical 
molecular diffusion of gaseous species is of the order of $10^{-5}[m^2/s]$, so the $D_b$ provides a useful estimate of the effective diffusion being larger than the pore scale induced effective diffusion.}. 
These considerations highlight the major role of the fibrous medium in the enhancement of mixing in liquids. 
Finally, the present findings on the effective diffusion can be directly applied to numerically solve  advection-dispersion-reaction macroscopic equations for the species 
flowing in real fibrous media, in order to design optimal electrodes.


%
%
%

\appendix*
\section{\label{app}Effect of the finite Reynolds number on the dispersion}
\begin{figure}[h]
\includegraphics[width=0.45\linewidth]{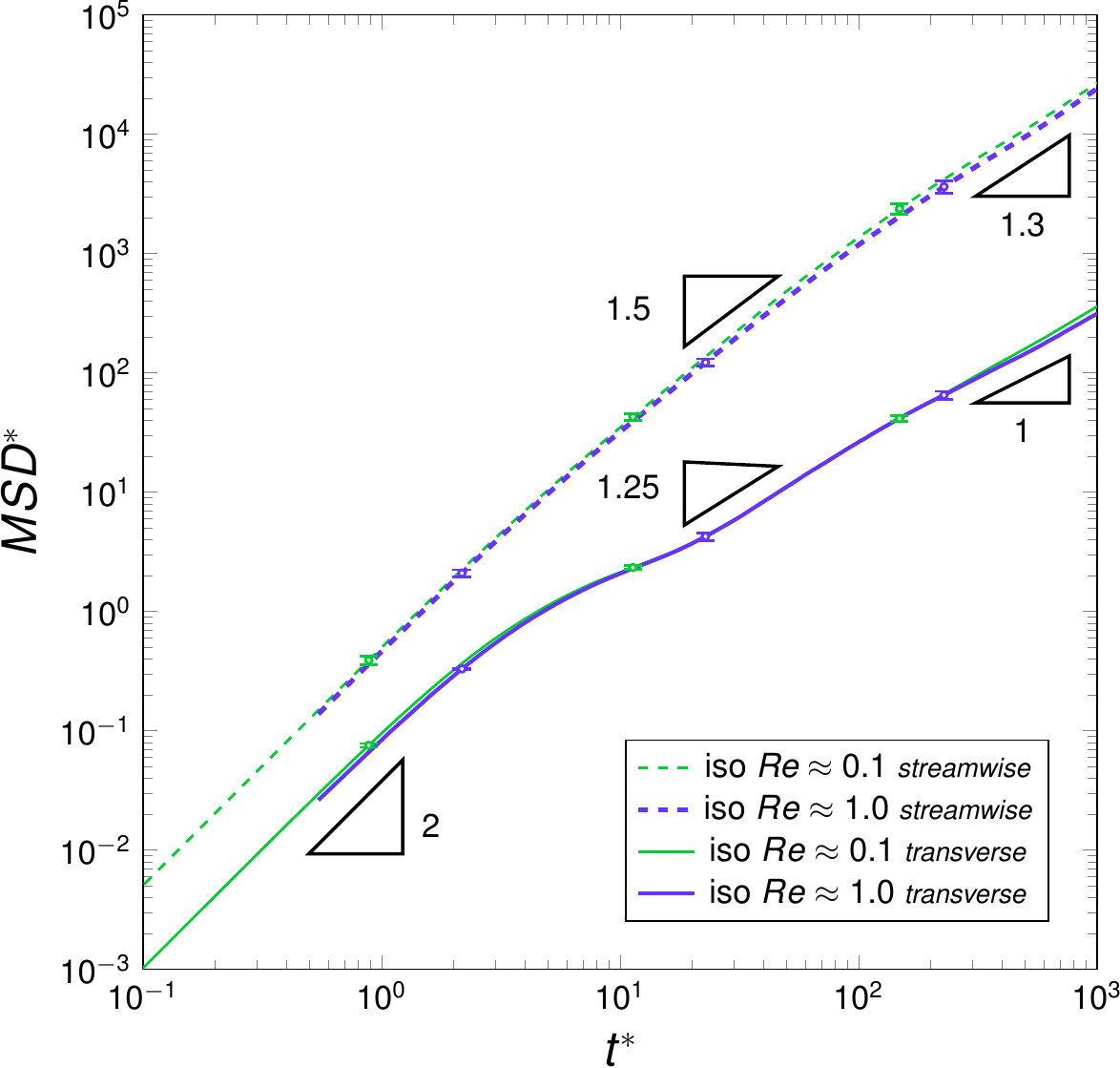}%
 \caption{\label{fig12}
 Mean Square Displacement with varying two different finite Reynolds numbers for isotropic fibrous media.}
\end{figure}

{In order to investigate the possible effects of inertia on dispersion dynamics at finite Reynolds number, simulations of flows through isotropic fibrous media have been performed at higher $\textit{Re}$ number}. Interestingly, no significant differences have been found by increasing the Reynolds number up to $\approx 1.0$ which can be considered an upper bound for Redox Flow Battery applications. Figure~\ref{fig12} shows a comparison between the Mean Square Displacements at $\textit{Re}\approx 0.1$ and at $\textit{Re}\approx1.0$, with the two curves overlapping almost perfectly. Since the Reynolds number range for the present application is $\textit{Re}=0.1\div1.0$, it can be concluded that the dispersion dynamics on flow batteries does not depend on the Reynolds number and consequently inertial effects are negligible.  

\begin{acknowledgments}
This work was supported as part of the MAESTRA project
(From Materials for Membrane-Electrode Assemblies to Electric Energy Conversion and Storage Devices, 2014-2016) funded by the University of Padua.
\end{acknowledgments}


\nocite{*}

\providecommand{\noopsort}[1]{}\providecommand{\singleletter}[1]{#1}%

\end{document}